\documentclass[journal,letterpaper,10pt]{IEEEtran}
\IEEEoverridecommandlockouts           

\usepackage{amsmath,amsfonts}
\usepackage{algorithmic}
\usepackage{algorithm}
\usepackage{array}
\usepackage[caption=false,font=normalsize,labelfont=sf,textfont=sf]{subfig}
\usepackage{textcomp}
\usepackage{stfloats}
\usepackage{url}
\usepackage{verbatim}
\usepackage{graphicx}
\usepackage{cite}

\usepackage{amsthm}
\usepackage[acronym]{glossaries}
\usepackage{booktabs}
\usepackage{xcolor}
\usepackage{amssymb}
\usepackage{threeparttable}
\usepackage{pifont}
    
\usepackage{threeparttable}
\usepackage{xcolor, colortbl}
\usepackage{tikz}             
\makeatletter
\@ifundefined{labelindent}{}{}
\makeatother
\usepackage{enumitem}
\usetikzlibrary{shapes,arrows,calc,positioning,decorations.pathmorphing,fit,backgrounds} 
\usepackage{pgfplots}
\usepackage{siunitx}
\usepackage[hidelinks]{hyperref}

\pgfplotsset{compat=newest}
\pgfplotsset{
	every axis/.append style={
		label style={font=\footnotesize},
		tick label style={font=\footnotesize},
		legend style={font=\footnotesize}
	}
}
\tikzset{
	-|-/.style={
		to path={
			(\tikztostart) -| ($(\tikztostart)!#1!(\tikztotarget)$) |- (\tikztotarget)
			\tikztonodes
		}
	},
	-|-/.default=0.5,
	|-|/.style={
		to path={
			(\tikztostart) |- ($(\tikztostart)!#1!(\tikztotarget)$) -| (\tikztotarget)
			\tikztonodes
		}
	},
	|-|/.default=0.5,
}
\tikzset{
	block/.style = {draw, rectangle,
		minimum height=0.9cm,
		align = center
	},
	input/.style = {coordinate,node distance=1cm},
	output/.style = {coordinate,node distance=0.8cm},
	arrow/.style={draw, -latex,node distance=2cm},
	pinstyle/.style = {pin edge={latex-, black,node distance=2cm}},
	sum/.style = {draw, circle, node distance=1cm},
	gain/.style = {
		regular polygon, regular polygon sides=3,
		draw, fill=white, text width=1em,
		inner sep=0mm, outer sep=0mm,
		shape border rotate=-90
	},
	dot/.style={circle,fill,draw,inner sep=0pt,minimum size=2pt}
}
\newtheoremstyle{normal} 
{}   
{}   
{}   
{}   
{\bfseries}  
{.}  
{ }  
{}   
\theoremstyle{normal}  
\newtheorem{assum}{Assumption}
\newtheorem{theorem}{Theorem}
\newtheorem{proposition}{Proposition}

\newacronym{AC}{AC}{Alternating Current}
\newacronym{BIBO}{BIBO}{Bounded-Input Bounded-Output}
\newacronym{DC}{DC}{Direct Current}
\newacronym{DFT}{DFT}{Discrete Fourier Transformation}
\newacronym{EV}{EV}{Electric Vehicle}
\newacronym{FOC}{FOC}{Field Oriented Controller}
\newacronym{HC}{HC}{Harmonic Control}
\newacronym{HSS}{HSS}{Harmonic Steady State}
\newacronym{LDV}{LDV}{Laser Doppler Vibrometer}
\newacronym{LTI}{LTI}{Linear Time Invariant}
\newacronym{LUT}{LUT}{Lookup Table}
\newacronym{MEMS}{MEMS}{Micro-Electro-Mechanical Systems}
\newacronym{MIMO}{MIMO}{Multiple-Input Multiple-Output}
\newacronym{NVH}{NVH}{Noise, Vibration, and Harshness}
\newacronym{PE}{PE}{Persistence of Excitation}
\newacronym{PWM}{PWM}{Pulse Width Modulation}
\newacronym[plural=PSMs]{PSM}{PSM}{Permanent-Magnet Synchronous Machine}
\newacronym{RLS}{RLS}{Recursive Least Squares}

\title{Adaptive Time-Domain Harmonic Control for Noise-Vibration-Harshness Reduction of Electric Drives}

\author{
	Klaus Herburger$^{1}$, Fabian Jakob$^{2}$, David G\"anzle$^{1}$, Maximilian Manderla$^{1}$, Andrea Iannelli$^{2}$%
	\thanks{$^{1}$ Corporate Research, Robert Bosch GmbH, 71272 Renningen, Germany. 
		klaus.herburger@de.bosch.com; david.gaenzle@de.bosch.com; maximilian.manderla@de.bosch.com}%
	\thanks{$^{2}$ University of Stuttgart, Institute for Systems Theory and Automatic Control, 70550 Stuttgart, Germany. 
		fabian.jakob@ist.uni-stuttgart.de; andrea.iannelli@ist.uni-stuttgart.de}%
	\thanks{Corresponding author: Klaus Herburger (klaus.herburger@de.bosch.com)}
}

\begin{document}

\maketitle
\thispagestyle{empty}
\pagestyle{empty}

\begin{abstract}

Reducing Noise, Vibration, and Harshness (NVH) in electric drives is crucial for applications such as electric vehicle drivetrains and heat-pump compressors, where strict NVH requirements directly affect user satisfaction and component longevity. This work presents the integration of an adaptive time-domain harmonic controller into an existing electric-drive control loop to attenuate harmonic disturbances. Three control structures are proposed and analyzed, along with a modified parameter-estimation scheme that reduces computational effort while preserving estimation accuracy, making the method suitable for embedded real-time implementation. To cope with fast operating-point changes, a delta-learning approach combines adaptive control with a lookup-table-based feedforward estimator, ensuring fast convergence and robustness. The proposed controller architectures are validated through simulation and testbench experiments on a permanent-magnet synchronous machine drive, demonstrating substantial NVH reductions across operating conditions. The results confirm that time-domain adaptive harmonic control offers a practical and theoretically grounded solution for real-time NVH mitigation in electric drives.

\end{abstract}

\section{INTRODUCTION}

Addressing \gls{NVH} in electric drives is crucial for the electrification of mobility and heating.
Beyond user comfort, mitigating \gls{NVH} also extends component lifetime.
At low speeds, electromagnetic \gls{NVH} dominates, which is caused by harmonic oscillations in forces and inverter voltages at multiples of the rotor’s electrical frequency \cite{psmNoise,husain,schroeder2015elektrische}. 

\gls{NVH} mitigation can be achieved either through constructive design or control-based methods \cite{langheckKit}. 
Inspired by harmonic disturbance rejection algorithms from other domains \cite{helicopterVibration,helicopterVibration2,noiseControl,gearboxVibration}, this work investigates a control-based approach: injecting sinusoidal control inputs at disturbance frequencies to cancel their effect. 
Implementation is challenged by the unknown current-to-\gls{NVH} transfer behavior and unmeasurable disturbances. 
However, surface accelerations, measurable e.g. by MEMS sensors, can be used as surrogate signals to quantify the NVH. They can therefore be utilized for offline or online estimation of the unknown transfer behavior.

An adaptive control approach is appealing in this context due to its ability to estimate system dynamics online. It maintains performance despite aging or temperature drift and reduces calibration effort, addressing key challenges in electric drives.
However, integrating adaptive controllers into electric drive control is non-trivial: the additional controller must coexist with existing ones without interference, be computationally efficient for embedded real-time execution, and converge rapidly to cope with changes in transfer behavior across operating points. 
This work addresses these challenges in the context of \gls{PSM} drives.

\subsection{Related work} \label{subsec:relatedwork}

Many methods for harmonic disturbance rejection in unknown systems have been proposed. 
If the open-loop dynamics between control input and measurement signal is asymptotically stable and partial model knowledge is available, adaptive feedforward cancellation can be applied \cite{adaptiveFeedforwardControl}. 
Feedback-based adaptive approaches are also possible but typically rely on model assumptions such as known relative degree or minimum-phase behavior \cite{feedbackAdaptiveControl}. 

A widely used alternative is \gls{HC}, developed independently across several research fields, which neglects transients and reduces the problem to disturbance attenuation in the \gls{HSS} \cite{hhc1,hhc2,hhc3,hhc4}. 
Classical HC requires sufficiently accurate phase knowledge of the transfer function to ensure stability, which is often difficult to obtain. 
To address this, adaptive HC methods have been proposed \cite{kamaldarDiss}. 
In frequency-domain adaptive HC, amplitudes and phases are extracted using a \gls{DFT} \cite{kamaldarFHC}. 
While effective, the large update period required for stability leads to slow convergence \cite{kamaldarDiss,helicopterVibration2}. 

Time-domain adaptive HC avoids this limitation by working directly with the measured signal. 
The stability was initially shown using averaging arguments \cite{pigg}; but also a more sophisticated Lyapunov-proof has arised \cite{kamaldarTHC}, which was used to demonstrate that larger adaptation gains can be used compared to frequency-domain adaptive HC. 

Building on these results, this work employs time-domain adaptive HC, as it achieves stability without model knowledge and offers faster convergence than frequency-domain methods.

\subsection{Objective and contribution}

This work investigates the integration of time-domain adaptive \gls{HC} into the control loop of an electric drive, with emphasis on a theoretically sound design that guarantees stability of both the adaptive controller and the overall system. The approach is validated through simulations and testbench experiments. 

To overcome the key challenges of controller coexistence, computational efficiency, and robustness under varying operating points, the main contributions are:
\begin{itemize}
	\item \textbf{Controller coexistence:} Development of three control structures that integrate adaptive time-domain \gls{HC} for \gls{NVH} reduction while ensuring compatibility with the existing control loop. 
	
	\item \textbf{Computational efficiency:} Modification of the estimation scheme to reduce matrix multiplications, achieving equal estimation accuracy with improved suitability for embedded real-time implementation. 
	
	\item \textbf{Robustness to operating changes:} Introduction of a delta learning approach that combines adaptive time-domain \gls{HC} with a \gls{LUT}-based feedforward estimator. An adaptive \gls{LUT} enables online updates that reflect parameter changes due to aging.
	
	\item \textbf{Experimental validation:} Implementation and testing of the proposed control structures on an e-drive testbench, confirming their feasibility and effectiveness under real operating conditions.
\end{itemize}

\subsection{Outline}

\autoref{sec:prelim} introduces preliminaries, before \autoref{chap:concept} presents the integration of the adaptive time-domain \gls{HC} and its parameter estimator, followed by a stability analysis. \autoref{sec:deltaLearning} introduces the delta learning approach. The concepts are validated through simulations (\autoref{sec:simResults}) and testbench experiments (\autoref{sec:testbenchResults}), before conclusions are drawn in \autoref{sec:conclusion}.
\section{Preliminaries} \label{sec:prelim}

\begin{figure}
	\centering
	\includegraphics[]{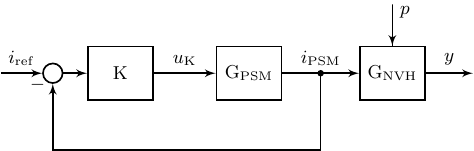}
%
%
%
%
%
%
	\caption{Control loop without HC.}
	\label{fig:controlWoHHC}
\end{figure}

The typical control loop of a \gls{PSM} is illustrated in \autoref{fig:controlWoHHC}. Here $i_{\mathrm{ref}}$ is a current that is typically pre-generated to match some reference torque $T_{\mathrm{ref}}$. The controller K regulates the currents by adjusting the voltages supplied to the \gls{PSM}, represented by the plant model $G_\mathrm{PSM}$ \cite{abuali}.
The \gls{NVH} emissions of the electric drive are represented by surface vibration quantities, denoted by $y$. They are modelled by an unknown, asymptotically stable \gls{LTI} transfer function $G_{\mathrm{NVH}}$ characterizing the dynamic behavior of the \gls{NVH} signal with input $i_{\mathrm{PSM}}$ and an additive external unmeasured output disturbance $p$ as
\begin{equation}
	y = G_{\mathrm{NVH}}i_{\mathrm{PSM}} + p.
\end{equation} 
Note that $i_\mathrm{PSM}$ is only indirectly controllable.

\subsection{Harmonic oscillations and NVH in electric drives} \label{subsec:nvhdrives}

Since $G_{\mathrm{NVH}}$ is asymptotically stable, the transient response $y_{\mathrm{trans}}$ of the measured \gls{NVH} emissions vanishes. Thus, the total response $y$ approaches its harmonic steady-state component $y_{\mathrm{hss}}$ asymptotically \cite{bernstein}, which is defined as
\begin{equation} \label{eq:yhssintro}
	\begin{aligned}
		y_{\mathrm{hss}}(t) &= \sum_\mathrm{i=1}^\mathrm{q} f_\mathrm{i}^\top(t)\,\theta_\mathrm{y,i}, \\
		f_\mathrm{i}(t) &= \begin{bmatrix} \sin(\omega_\mathrm{i} t) \\ \cos(\omega_\mathrm{i} t) \end{bmatrix}, &
		\theta_\mathrm{y,i} &= \begin{bmatrix} \theta_\mathrm{y,s,i} \\ \theta_\mathrm{y,c,i} \end{bmatrix} \in \mathbb{R}^2.
	\end{aligned}
\end{equation}
The goal is to achieve $y_{\mathrm{hss}}(t) = 0$.
In~\eqref{eq:yhssintro}, $\omega_\mathrm{i} > 0$ are the disturbance frequencies, which must be known to apply harmonic control.
\gls{NVH} in electric machines are typically caused by harmonic oscillations in the electromagnetic forces and inverter voltages \cite{husain}\cite{schroeder2015elektrische}. Crucially, empirical findings show that those oscillations occur at frequencies $\omega_\mathrm{el,m} = \mathrm{m} \omega_\mathrm{el}$ that are multiple integers of the known electrical frequency $\omega_\mathrm{el}$ \cite{husain}\cite{schroeder2015elektrische}. 
The problem of mitigating harmonics at different frequencies can be considered decoupled from each other, therefore we refer to any $\omega_\mathrm{el,m}$ as just $\omega_\mathrm{i}$ in the following.
Moreover, considering currents in the dq-coordinate system, it has been observed that \gls{NVH} is primarily induced by the $q$-component $i_\mathrm{q}$ \cite{harries}.
Thus, for the remainder of this work, we consider one-dimensional control inputs. Extensions to multidimensional signals exist and are straightforward, which may be explored when two independent measurements $y_\mathrm{1}$ and $y_\mathrm{2}$ are available and if both the d and q component shall be exploited \cite{kamaldarTHC}.

\subsection{Harmonic control} 

In \gls{HC} it is assumed to have access to a control input $u$ and a measureable performance signal $y$ for which a dynamic relation 
\begin{equation}  \label{eq:system}
	y(t) = G_{\mathrm{u \rightarrow y}}(s)[u(t)]+p(t),
\end{equation}
with unknown disturbance $p$ holds.
Crucially, all signals are assumed to only consist of their HSS response, i.e., the transient is neglected and
\begin{equation}
	p(t) = \sum_\mathrm{i=1}^\mathrm{q}f_\mathrm{i}^\mathrm{T}(t)\theta_\mathrm{p,i} \qquad u(t) = \sum_\mathrm{i=1}^\mathrm{q}f_\mathrm{i}(t)^\mathrm{T}\theta_\mathrm{u,i}.
\end{equation}
Using linear system theory \cite{bernstein}, it can be shown that the \gls{HSS} response of~\eqref{eq:system} results in
\begin{equation} \label{eq:yhssorigi}
	y(t) = \sum_\mathrm{i=1}^\mathrm{q}f_\mathrm{i}^\mathrm{T}(t)(G_\mathrm{i}\theta_\mathrm{u,i}+\theta_\mathrm{p,i})
\end{equation}
with
\begin{equation} \label{eq:Hfirst}
	G_\mathrm{i} = \begin{bmatrix}
		\text{Re} \ G_{\mathrm{u \rightarrow y}}(j\omega_\mathrm{i}) & -\text{Im} \ G_{\mathrm{u \rightarrow y}}(j\omega_\mathrm{i}) \\
		\text{Im} \ G_{\mathrm{u \rightarrow y}}(j\omega_\mathrm{i}) & \text{Re} \ G_{\mathrm{u \rightarrow y}}(j\omega_\mathrm{i}) \\
	\end{bmatrix} \in \mathbb{R}^{2 \times 2}.
\end{equation}
The goal is to supress harmonic steady state oscillations, and if $G_\mathrm{i}$ and $\theta_\mathrm{p,i}$ are known, $\theta_\mathrm{u,i}$ can be calculated as
\begin{equation} \label{eq:calcu}
	\theta_\mathrm{u,i} = -G_\mathrm{i}^{-1}\theta_\mathrm{p,i}
\end{equation} 
to achieve $y_{\mathrm{hss}} = 0$, making the approach both intuitive and straightforward. In our setup however, the main challenge is to estimate the unknown transfer function $G_\mathrm{i}$ and the unknown disturbance $\theta_\mathrm{p,i}$ at the relevant frequencies. 
Since $G_\mathrm{i}$ and $\theta_\mathrm{p,i}$ change online due to varying operating points and aging effects, it is desirable to estimate them in real-time, yielding an adaptive control scheme. 
Note that adaptive \gls{HC} approaches have already been studied — see, e.g., \cite{kamaldarFHC,kamaldarTHC,pigg}, where the transfer matrix $G_\mathrm{i}$ is estimated online. Building on these preliminaries, the following section discusses how the adaptive time-domain \gls{HC} can be integrated into the control loop of the electric drive.
\section{Adaptive time-domain hc in electric drives: Control structures, estimation, and stability} \label{chap:concept}

This section develops the core contribution of this work: the integration of adaptive time-domain \gls{HC} into the control structure of electric drives to reduce \gls{NVH} emissions. 

\subsection{Integration into the control loop} \label{sec:integration}

The objective is to incorporate an adaptive time-domain \gls{HC} into the control loop to mitigate \gls{NVH} emissions without affecting the existing controller K. The input to the unknown system $i_\mathrm{PSM}$ cannot be manipulated directly, but only through the control voltages $u_\mathrm{K}$.
Three control structures are proposed, where each structure offers unique advantages/disadvantages and may be favored based on the application and on which parts of the control loop may be accessible.
In all control structures, the system formulation~\eqref{eq:system} 
is used, but $u$ and $G_\mathrm{u \rightarrow y}$ vary depending on the structure.

\begin{figure}[t]
	\centering
	\subfloat[\label{fig:Structure1}]{
		\includegraphics[]{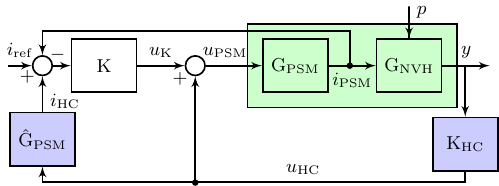}
}\hfil
	\subfloat[\label{fig:Structure2}]{
		\includegraphics[]{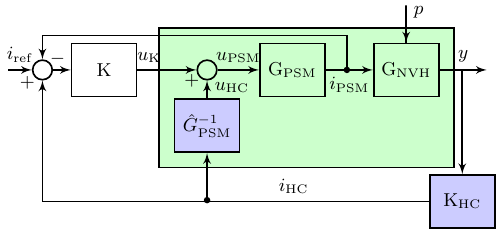}
}\hfil
	\subfloat[\label{fig:Structure3}]{
		\includegraphics[]{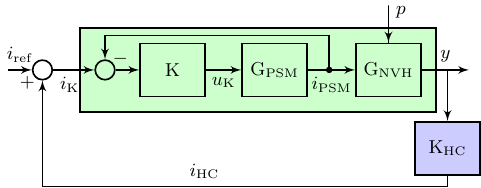}
}\caption{Control architectures considered in this work. 
		The shaded area indicates the part of the system assumed unknown. (a) Structure 1 -- Acceleration to voltage. (b) Structure 2 -- Acceleration to measured current. (c) Structure 3 -- Acceleration to reference current.}
	\label{fig:ControlStructures}
\end{figure}

\subsubsection{Structure 1 -- Acceleration to voltage} \label{subsec:struc1}

In Structure 1 the control output is the voltage $u_\mathrm{HC}$ that is added to $u_\mathrm{K}$ to get $u_\mathrm{PSM}$, as illustrated in Figure\autoref{fig:Structure1}.
The unknown system results in
\begin{equation}
	y = G_\mathrm{u_\mathrm{PSM} \rightarrow y}u_\mathrm{PSM} + p,
\end{equation}
i.e., $u=u_\mathrm{PSM}$ and $G_\mathrm{u\rightarrow y}=G_\mathrm{u_\mathrm{PSM}\rightarrow y} = G_\mathrm{i_\mathrm{PSM}\rightarrow y}G_\mathrm{PSM}$. This approach has the advantage that frequency-domain \gls{HC} concepts as in \cite{kamaldarFHC} can be implemented with the same control structure.
The disadvantage is that the known model $G_\mathrm{PSM}$ is not leveraged in the \gls{HC}, since it treats this part of the system as unknown. 
Crucially, a decoupling block $\hat{G}_\mathrm{PSM}$ avoids that K and \gls{HC} influence each other due to their conflicting control goals. 
Specifically, 
\begin{equation}
	i_\mathrm{HC} = \hat{G}_\mathrm{PSM} u_\mathrm{HC}
\end{equation}
ensures that the transfer behavior from $ i_\mathrm{ref}$ to $i_\mathrm{PSM}$ remains the closed loop transfer behavior of the nominal control structure. If $\hat{G}_\mathrm{PSM}$ is equal to $G_\mathrm{PSM}$, then
\begin{equation}
	\begin{aligned}
		i_\mathrm{K} &= i_\mathrm{ref} - i_\mathrm{PSM} + i_\mathrm{HC} \\
		&= i_\mathrm{ref} - G_\mathrm{PSM}(u_\mathrm{K} + u_\mathrm{HC}) + \hat{G}_\mathrm{PSM} u_\mathrm{HC} \\
		&= (1+G_\mathrm{PSM}K)^{-1}i_\mathrm{ref},
		\label{eq:Dec1}
	\end{aligned}
\end{equation}
which corresponds exactly to the nominal control structure.
Accurate models are reasonable assumptions in e-drives, however, care must be taken for model mismatches regarding the feedforward prediction.

\subsubsection{Structure 2 -- Acceleration to measured current} \label{subsec:struc2}

Structure 2 aims at estimating the transfer behavior from $i_\mathrm{PSM}$ to $y$. 
Thus, the output of the controller is the current that would need to be added to $i_\mathrm{PSM}$. 
Since it is not possible to control $i_\mathrm{PSM}$ directly, the control output $i_\mathrm{HC}$ is converted to $u_\mathrm{HC}$ using an inverse model of the \gls{PSM} that is denoted as $\hat{G}_\mathrm{PSM}^\mathrm{-1}$. The control loop for Structure 2 is shown in Figure\autoref{fig:Structure2}.
It considers
\begin{equation}
	\begin{aligned}
		y &= G_\mathrm{i_\mathrm{HC} \rightarrow y}i_\mathrm{HC} + p \\
		&= G_\mathrm{i_\mathrm{PSM} \rightarrow y}G_\mathrm{PSM}\hat{G}_\mathrm{PSM}^\mathrm{-1}i_\mathrm{HC} + p
	\end{aligned}
\end{equation}
as the unknown system that is to be estimated.
To implement $\hat{G}_\mathrm{PSM}^\mathrm{-1}$, a time delay is introduced at the input to avoid acausality in the inverse \gls{PSM} model.
If the condition $G_\mathrm{PSM}\hat{G}_\mathrm{PSM}^\mathrm{-1} = 1$
holds, the \gls{HC} does not influence K, if $i_\mathrm{HC}$ is subtracted from $i_\mathrm{PSM}$. Then, only the current resulting from $u_\mathrm{K}$ remains in the feedback loop of K and~\eqref{eq:Dec1} holds.
The main advantage of Structure 2 over Structure 1 is then that the \gls{PSM} is not part of the estimated system.

\subsubsection{Structure 3 -- Acceleration to reference current}

Another way to achieve that the \gls{PSM} is not part of the estimated system is to add the control output $i_\mathrm{HC}$ to the reference current $i_\mathrm{ref}$ to get $i_\mathrm{K}$.
Structure 3 does -- in contrast to the other two structures -- not need an exact model of $G_\mathrm{PSM}$ to achieve the decoupling of the controllers. The resulting control structure is shown in Figure\autoref{fig:Structure3}.
This results in the following representation of the unknown and to be estimated system:
\begin{equation}
	y = G_\mathrm{i_\mathrm{K} \rightarrow y}i_\mathrm{K} + p.
\end{equation}
Since 
\begin{equation}
	i_\mathrm{PSM} = \frac{G_\mathrm{PSM}K}{1+G_\mathrm{PSM}K} i_\mathrm{K},
\end{equation} the (known) closed loop of K is part of the estimated system with 
\begin{equation}
	y = G_\mathrm{i_\mathrm{PSM} \rightarrow y}  \frac{G_\mathrm{PSM}K}{1+G_\mathrm{PSM}K} i_\mathrm{K} + p,
\end{equation}
which is a disadvantage of this structure. 

\subsubsection{Comparison of the three control structures}

\begin{table}
	\centering
	\caption{Overview of the three control structures.}
	\begin{tabular}{@{}>{\raggedright}p{1.3cm} p{0.8cm} p{2.5cm} p{2.3cm}@{}}
		\toprule
		\textbf{Method} & \textbf{Input} & \textbf{Main advantage} & \textbf{Main disadvantage} \\ 
		\midrule
		Structure 1 & $u_\mathrm{PSM}$ & Similar to frequency-domain structure & \gls{PSM} model is assumed unknown \\ 
		Structure 2 & $i_\mathrm{PSM}$ & System assumed unknown equals the unknown system & \gls{PSM} model needs to be known exactly for full decoupling \\ 
		Structure 3 & $i_\mathrm{K}$ & No controller decoupling necessary & K closed loop is assumed unknown \\ 
		\bottomrule
	\end{tabular}
	\label{tab:controllerStructures}
\end{table}

In this section, three control structures are presented, that are designed to integrate adaptive time-domain \gls{HC} techniques into the existing control loop. An overview is shown in \autoref{tab:controllerStructures}. It highlights the input $u$ to the estimated system, the main advantage and the main disadvantage of each structure.

\subsection{Online parameter estimation} \label{sec:setup}

\begin{figure*}
	\centering
	\includegraphics[]{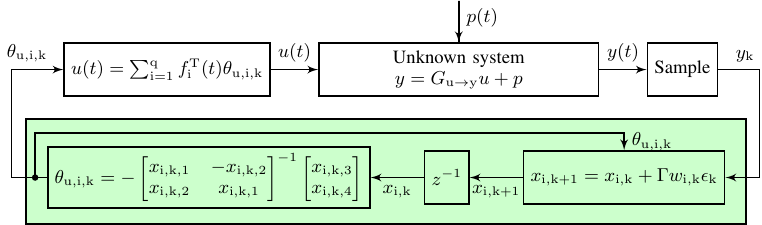}
	\caption{Feedback interconnection with adaptive time-domain harmonic controller (inside shaded area).}
	\label{fig:adaptiveController}
\end{figure*}

Since the true parameters $G_\mathrm{i}^*$ and $\theta_\mathrm{p,i}^*$ are unknown, the controller relies on their estimates $G_\mathrm{i}$ and $\theta_\mathrm{p,i}$. 
An estimation scheme is proposed for the general formulation~\eqref{eq:yhssorigi}, thus applicable to all three control structures, and with the goal of identifying $G_\mathrm{i}$ and $\theta_\mathrm{p,i}$ online.
Define the sampled data as
\begin{align}
	y_\mathrm{k} = y(kt_\mathrm{s}), \qquad
	f_\mathrm{i,k} = f_\mathrm{i}(kt_\mathrm{s})
\end{align} 
for each $k \in \mathbb{N}$ with the sampling time $t_\mathrm{s}$.
Using the approximation
\begin{equation} \label{eq:ydiscrete}
	y(kt_\mathrm{s}) = y_{\mathrm{hss}}(kt_\mathrm{s}, \theta_\mathrm{u,1,k}, \dots , \theta_\mathrm{u,q,k}),
\end{equation}
$y_\mathrm{k}$ can be expressed as
\begin{equation} \label{eq:ydisc}
	y_\mathrm{k} = \sum_\mathrm{i=1}^\mathrm{q} f_\mathrm{i,k}^\mathrm{T} \left(G_\mathrm{i}^*\theta_\mathrm{u,i,k} +\theta_\mathrm{p,i}^* \right).
\end{equation}
Analogously, the estimation $\hat{y}_\mathrm{k}$ is defined as
\begin{equation} \label{eq:yhat}
	\hat{y}_\mathrm{k} = \sum_\mathrm{i=1}^\mathrm{q} f_\mathrm{i,k}^\mathrm{T} \left(G_\mathrm{i,k} \theta_\mathrm{u,i,k}+\theta_\mathrm{p,i,k}\right).
\end{equation}
More compactly, we write~\eqref{eq:yhat} as
\begin{equation} \label{eq:yestimation}
	\hat{y}_\mathrm{k} = \sum_\mathrm{i=1}^\mathrm{q} w_\mathrm{i,k}^\mathrm{T}x_\mathrm{i,k},
\end{equation}
with the unknown parameter vector 
\begin{equation} \label{eq:estiparameters1}
	x_\mathrm{i,k}^\mathrm{T} := \begin{bmatrix}
		\text{Re} \ G_{\mathrm{u \rightarrow y}}(j\omega_\mathrm{i})_\mathrm{k} & \text{Im} \ G_{\mathrm{u \rightarrow y}}(j\omega_\mathrm{i})_\mathrm{k} & \theta_\mathrm{p,s,i,k} & \theta_\mathrm{p,c,i,k} \end{bmatrix}
\end{equation}
and the regressor
\begin{equation} \label{eq:estiparameters2}
	w_\mathrm{i,k} := \begin{bmatrix}
		\theta_\mathrm{u,s,i,k}\sin(\omega_\mathrm{i} kt_\mathrm{s})+\theta_\mathrm{u,c,i,k}\cos(\omega_\mathrm{i} kt_\mathrm{s}) \\ \theta_\mathrm{u,s,i,k}\cos(\omega_\mathrm{i} kt_\mathrm{s})-\theta_\mathrm{u,c,i,k}\sin(\omega_\mathrm{i} kt_\mathrm{s}) \\ \sin(\omega_\mathrm{i} kt_\mathrm{s}) \\ \cos(\omega_\mathrm{i} kt_\mathrm{s})
	\end{bmatrix}.
\end{equation}
Here, the subscripts “s” and “c” denote the sine and cosine components of the corresponding parameters 
$\theta_{\mathrm{u}}$ and $\theta_{\mathrm{p}}$, respectively, as introduced in~\eqref{eq:yhssintro}.
We define the cost function
\begin{equation} \label{eq:paramestij}
	J_\mathrm{k}(x_\mathrm{k}) = \frac{\eta_\mathrm{k}\|y_\mathrm{k} -  \sum_\mathrm{i=1}^\mathrm{q} w_\mathrm{i,k}^\mathrm{T}x_\mathrm{i,k}\|^2}{2} = \frac{\| \epsilon_\mathrm{k}\|^2}{2 \eta_\mathrm{k}}
\end{equation}
which is to be minimized, where the normalization error is defined as
\begin{equation} \label{eq:paramestieps}
	\epsilon_\mathrm{k} = \eta_\mathrm{k}(y_\mathrm{k}-\hat{y}_\mathrm{k}).
\end{equation}
Since this a positive semi-definite quadratic cost function, standard methods with provable convergence rates can be employed. Considering the computational budget, the gradient method
is used.
Following \cite{robustAdaptiveControl}, this leads to the adaptive law 
\begin{equation} \label{eq:al}
	x_\mathrm{i,k+1} = x_\mathrm{i,k} + \Gamma w_\mathrm{i,k} \epsilon_\mathrm{k}.
\end{equation}
Here specifically, the normalization factor is chosen as
\begin{equation} \label{eq:normalization}
	\eta_\mathrm{k} = \frac{1}{1+\sum_\mathrm{i=1}^\mathrm{q} \|\theta_\mathrm{u,i,k}\|^2}.
\end{equation}
This particular choice of the normalization factor is needed for the stability theorem in the next section and is one modification over \cite{pigg}.
The learning rate $\Gamma$ is chosen as 
\begin{equation}
	\Gamma = \text{diag}(\gamma_\mathrm{G}, \gamma_\mathrm{G}, \gamma_\mathrm{p}, \gamma_\mathrm{p})
\end{equation}
to provide individual learning rates $\gamma_\mathrm{G}, \gamma_\mathrm{p} > 0$ for the transfer function and the disturbance parameters. 
Note that unlike in \cite{kamaldarTHC}, $G_\mathrm{i,k}$ and $\theta_\mathrm{p,i,k}$ are estimated jointly, which will be especially beneficial in the context of computational limitations.
Finally, the estimated parameters $x_\mathrm{i,k}$ are then used to construct $G_\mathrm{i,k}$ and $\theta_\mathrm{p,i,k}$.
This approach ensures that the block structure of $G_\mathrm{i,k}$ is preserved without adding additional terms to the estimation equations as in \cite{kamaldarTHC}. 
In summary, the control law is given by~\eqref{eq:al} with
\begin{equation} \label{eq:control}
	\theta_\mathrm{u,i,k} = - G_\mathrm{i,k}^{-1} \theta_\mathrm{p,i,k}.
\end{equation}
The control architecture is shown in \autoref{fig:adaptiveController}. 

\subsection{Stability theorem} \label{sec:stabanalysis}

Since the control law~\eqref{eq:control} is used for all $k \in \mathbb{N}$, it is also used when the parameters are incorrectly initialized. A stability theorem ensures that stability of the controller and convergence of $y_\mathrm{k}$ to zero can be guaranteed for all initial conditions $G_\mathrm{i,0}$ and $\theta_\mathrm{p,i,0}$, i.e. the same properties as in \cite{kamaldarTHC} can be shown for the computationally efficient estimation scheme proposed in the previous subsection.

We make the following assumptions:
\begin{assum} \label{assum:standing}
	The following conditions hold for all $i = 1,\dots,\mathrm{q}$ and all $k \in \mathbb{N}$:
	\begin{enumerate}[label=(1\alph*)]
		\item \label{assum:standing:a} The transfer function satisfies 
		$\mathrm{rank}\, G_{\mathrm{u \rightarrow y}}(j\omega_\mathrm{i}) = l$.
		\item \label{assum:standing:b} The disturbance frequencies $\omega_1,\dots,\omega_\mathrm{q}$ are known.
		\item \label{assum:standing:c} The output satisfies 
		$y_\mathrm{k} = y_{\mathrm{hss}}(kt_\mathrm{s},\theta_{\mathrm{u,1,k}},\dots,\theta_{\mathrm{u,q,k}})$.
		\item \label{assum:standing:d} There exists $\epsilon > 0$ such that 
		$\lambda_{\min}(G_\mathrm{i,k}G_\mathrm{i,k}^\top) > \epsilon$.
	\end{enumerate}
\end{assum}
\noindent Assumption~\ref{assum:standing:a} implies that the number of actuators must be at least as large as the number of performance measurements. 
This is always fulfilled in the considered setup, since $y \in \mathbb{R}$ and $u \in \mathbb{R}$.
Assumption~\ref{assum:standing:b} is fulfilled, since the harmonic disturbance frequencies are known to be integer multiples of the known electrical frequency $\omega_\mathrm{el}$. 
The \gls{HSS} assumption~\ref{assum:standing:c} indicates the stability theorem is valid assuming the transient response $y_\mathrm{trans}$ vanished. Given that $G_\mathrm{NVH}$ is asymptotically stable with relatively fast dynamics, this assumption is considered to be met..
Assumption~\ref{assum:standing:d} infers that there exists a singularity at $(x_\mathrm{i,k,1}, x_\mathrm{i,k,2}) = 0$ that is to be avoided. Using
\begin{equation}
	G_\mathrm{i,k} = \begin{bmatrix}
		x_\mathrm{i,k,1} & -x_\mathrm{i,k,2} \\
		x_\mathrm{i,k,2} & x_\mathrm{i,k,1} \\
	\end{bmatrix} \in \mathbb{R}^{2 \times 2}
\end{equation}
with $x_\mathrm{i,k,1} = \text{Re} \ G_{\mathrm{u \rightarrow y}}(j\omega_\mathrm{i})_\mathrm{k}$ and $x_\mathrm{i,k,2} = \text{Im} \ G_{\mathrm{u \rightarrow y}}(j\omega_\mathrm{i})_\mathrm{k}$, it follows 
\begin{equation}
	G_\mathrm{i,k}G_\mathrm{i,k}^\mathrm{T} = \begin{bmatrix}
		x_\mathrm{i,k,1}^2 + x_\mathrm{i,k,2}^2 & 0\\
		0 & x_\mathrm{i,k,1}^2 + x_\mathrm{i,k,2}^2 \\
	\end{bmatrix} 
\end{equation}
and therefore 
\begin{equation}
	\lambda_\mathrm{min}(G_\mathrm{i,k}G_\mathrm{i,k}^\mathrm{T}) = x_\mathrm{i,k,1}^2 + x_\mathrm{i,k,2}^2.
\end{equation} 
This implies $\lambda_\mathrm{min}(G_\mathrm{i,k}G_\mathrm{i,k}^\mathrm{T}) > \epsilon$ if $(x_\mathrm{i,k,1}, x_\mathrm{i,k,2}) \neq 0$. 
This requirement is inherently met in the initial estimation phase if the initial values $x_\mathrm{i,0,1}, x_\mathrm{i,0,2}$ are chosen sufficiently large. After the initial phase the singularity can be avoided by suitable algorithm modifications, such as the addition of an arbitrary perturbation to the parameters in regions of small amplitudes or the interruption of the $x_\mathrm{i,k,1}, x_\mathrm{i,k,2}$ update until the gradient forces an amplitude growth again.

For the stability theorem, define the closed-loop system 
\begin{equation} \label{closedloop}
	y_\mathrm{k} = \sum_\mathrm{i=1}^\mathrm{q} f_\mathrm{i,k}^\mathrm{T} \left(\theta_\mathrm{p,i}^* - G_\mathrm{i}^* G_\mathrm{i,k}^\mathrm{-1}\theta_\mathrm{p,i,k}\right),
\end{equation}
which follows when the control law~\eqref{eq:control} is plugged into the system formulation~\eqref{eq:ydisc}.
Define the parameter estimation error as 
\begin{equation} \label{eq:estierror}
	\tilde{x}_\mathrm{i,k} = x_\mathrm{i,k} - \alpha x_\mathrm{i}^*.
\end{equation}
The following theorem provides the stability conditions for the closed-loop system.
\begin{theorem} \label{stabilityproof}
	Consider the closed-loop system~\eqref{closedloop} that is generated using the estimation~\eqref{eq:al} and the control~\eqref{eq:control} and consider the parameter estimation error~\eqref{eq:estierror}. 
	Assume that Assumption \autoref{assum:standing} is satisfied. Then, for all \( \alpha > \frac{\gamma_\mathrm{G}}{2} + q \gamma_\mathrm{p} \), the following statements hold:
	\begin{enumerate}
		\item \( (\tilde{x}_\mathrm{1,k} \dots \tilde{x}_\mathrm{q,k}) \equiv 0 \) is a uniformly Lyapunov stable equilibrium of~\eqref{closedloop} and~\eqref{eq:estierror}.
		
		\item Let \( \tilde{x}_\mathrm{1,0} \dots \tilde{x}_\mathrm{q,0} \in \mathbb{R}^4 \). Then \( x_\mathrm{i,k} \) and \( \theta_\mathrm{u,i,k} \) are bounded and \( \lim_\mathrm{k \to \infty} y_\mathrm{k} = 0 \).
	\end{enumerate}
\end{theorem}
\noindent The proof follows along the lines of~\cite{kamaldarTHC}, with necessary adaptations for the 
modified estimator scheme. A concise proof sketch is given in Appendix~\ref{proof}.
\noindent Note that the proof does not introduce any restrictions for $\tilde{x}_\mathrm{1,0} \dots \tilde{x}_\mathrm{q,0}$. Therefore, part 2) of \autoref{stabilityproof} holds globally, which means that 
\begin{enumerate}
	\item $y_\mathrm{k}$ converges to zero for all initial parameter estimates ${x}_\mathrm{1,0} \dots {x}_\mathrm{q,0}$. 
	\item It is always possible to find $\alpha$ such that $\alpha > \frac{\gamma_\mathrm{G}}{2} + q \gamma_\mathrm{p}$ holds and~\eqref{eq:estierror} is fulfilled.
\end{enumerate}

\subsection{Parameter convergence} \label{sec:parameterconv}

Convergence of $x_\mathrm{i,k}$ to the true parameters $x_\mathrm{i}^*$ is neither required by the stability proof, nor is it implied by it. However, studying when these parameters converge to their actual true value will be beneficial for the modifications that are proposed in the following sections. Consider the following Proposition.

\begin{proposition} \label{prop:rhok} (\cite{nesterov}) 
	Consider the harmonic mode $i$ and assume $J_\mathrm{k}$ is $\mu_\mathrm{i,k}$-strongly convex and $L_\mathrm{i,k}$-smooth in $x_\mathrm{i}$. 
	Then, if the learning rate is chosen as
	\begin{equation*}
		\gamma_\mathrm{G}, \gamma_\mathrm{p} \leq \frac{2}{\mu_\mathrm{k}+L_\mathrm{k}}
	\end{equation*} for all $k$, then the adaptive law~\eqref{eq:al} generates a sequence ${x_\mathrm{i,k}}$ such that
	\begin{equation*} \label{eq:parameterconv}
		\|  x_\mathrm{i,k} - x_\mathrm{i}^* \| \leq \rho_\mathrm{i,k}^\mathrm{k} \| x_\mathrm{i,0} - x_\mathrm{i}^* \|
	\end{equation*}
	with
	\begin{equation*}\label{eq:rhok}
		\rho_\mathrm{i,k} = 1 - \frac{2 \min(\gamma_\mathrm{G}, \gamma_\mathrm{p}) \mu_\mathrm{i,k} L_\mathrm{i,k}}{\mu_\mathrm{i,k} + L_\mathrm{i,k}}.
	\end{equation*}
\end{proposition}
\noindent Since $J_\mathrm{k}$ is twice differentiable, $\mu_\mathrm{i,k}$ and $L_\mathrm{i,k}$ can be determined as minimum and maximum eigenvalue of the Hessian $H_\mathrm{i,k}$ of $J_\mathrm{k}$.
However, note that if $J_\mathrm{k}$ is defined as in~\eqref{eq:paramestij}, then 
\begin{equation} \label{eq:hessian}
	H_\mathrm{i,k} = \frac{\partial^2 J}{\partial x_\mathrm{i,k}^2} = \eta_\mathrm{k}\,w_\mathrm{i,k}^\top w_\mathrm{i,k},
\end{equation}
which is rank deficient. Thus, \autoref{prop:rhok} is not applicable to~\eqref{eq:paramestij} and no statement about parameter convergence can be made.

To circumvent this issue, dynamic regressor extension can be applied \cite{drem2,drem1}. 
Note that to uniquely identify four parameters in a regression problem, at least four linearly independent measurement equations are required. When having only one measurement at each $k$, a widely applied technique is to filter this sequence with four linearly independent linear filters, thus giving four linearly independent sequences which yield a well-posed problem. Consider 
\begin{equation}
	\mathcal{H}(z) = \left[1 \ \mathcal{H_\mathrm{1}}(z) \ \mathcal{H_\mathrm{2}}(z) \ \mathcal{H_\mathrm{3}}(z)\right].
\end{equation}
with 
\begin{equation}
	\mathcal{H_\mathrm{l}}(z) = \frac{(1-T_\mathrm{l})z}{z-T_\mathrm{l}}.
\end{equation}
Distinct values of the time-constants $T_\mathrm{l}$ are chosen to guarantee linear independence among the generated filtered signals.
Define
\begin{equation}
	\psi_\mathrm{k} := \mathcal{H}[y_\mathrm{k}] \in \mathbb{R}^\mathrm{4}, \quad \Phi_\mathrm{i,k} := \mathcal{H}[w_\mathrm{i,k}]\in \mathbb{R}^\mathrm{4 \times 4}
\end{equation}
and
\begin{equation} \label{eq:dynparamestij}
	J_\mathrm{k}(x_\mathrm{k}) = \frac{\eta_\mathrm{k}\|\psi_\mathrm{k} -  \sum_\mathrm{i=1}^\mathrm{q} \Phi_\mathrm{i,k}^\mathrm{T}x_\mathrm{i,k}\|^2}{2}.
\end{equation}
Then, 
\begin{equation}
	H_\mathrm{i,k} = \eta_\mathrm{k} \Phi_\mathrm{i,k}^\top \Phi_\mathrm{i,k},
\end{equation}
and thus, regularity of the Hessian matrix can be achieved. 

Note that $H_\mathrm{i,k}$ may still become rank deficient, for instance when $w_\mathrm{i,k} \equiv 0$. Hence, the data must still satisfy some excitation requirements. In contrast, if $H_\mathrm{i,k}$ has full rank and hence, $\rho_\mathrm{i,k} < 1$, $w_\mathrm{i,k}$ must have been sufficiently rich. Therefore, a phase of iterations where $\rho_\mathrm{i,k} < 1$ indicates a phase of qualitative parameter updates and therefore, a good estimation phase. This intuition will be significant for later proposed modifications of the estimation scheme.
\section{Delta learning based feedforward estimation} \label{sec:deltaLearning}

In non-adaptive \gls{HC}, it is well known that the phase of the model used for the control law 
$\angle G_\mathrm{u\rightarrow y}(\omega_\mathrm{i})$ and the phase of the true system $\angle G_\mathrm{u\rightarrow y}(\omega_\mathrm{i})^*$ must lie within a $\ang{90}$ margin to guarantee stability \cite{pigg}. 
In contrast, the adaptive scheme ensures stability for all parameter initializations, even when they are arbitrary bad. However, while preserving stability, deviations from the $\pm \ang{90}$ range may still degrade transient performance drastically.
This effect is particularly relevant during speed changes, where $\angle G_\mathrm{u \rightarrow y}(j \omega_\mathrm{i})$ shifts due to its frequency dependence. The stable range shifts accordingly, requiring time for the estimated parameters to adapt. 
To counteract this effect, this section introduces a delta learning approach that integrates a \gls{LUT}-based feedforward estimation. 
An adaptive \gls{LUT} is proposed that continuously updates during operation and reuses stored parameter values at previously visited operating points.

\subsection{Integration of delta learning into the harmonic controller} \label{subsec:integrationDelta}

\begin{figure}
	\centering
	\includegraphics[]{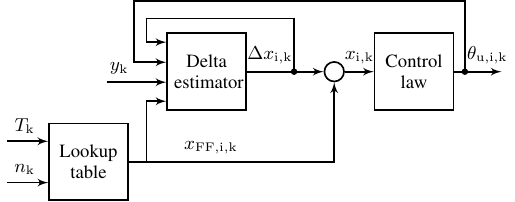}
	\caption{Delta learning adaptive controller.}
	\label{fig:deltaadaptivecontroller}
\end{figure}

Recall that the adaptive controller consists of an estimator that calculates $x_\mathrm{i,k+1}$ based on $y_\mathrm{k}$ and the previous output of the controller $\theta_\mathrm{u,i,k}$, along with a control law that computes $\theta_\mathrm{u,i,k+1}$ based on $x_\mathrm{i,k+1}$. \autoref{fig:deltaadaptivecontroller} shows how this adaptive controller is extended to incorporate the delta learning concept.
The \gls{LUT} provides feedforward parameter estimates $x_\mathrm{FF,i,k}$ based on the current operating point, which is defined by the torque $T_\mathrm{k}$ and the rotor speed $n_\mathrm{k}$. 

Instead of using~\eqref{eq:paramestij}, we consider the shifted optimization problem
\begin{equation} \label{eq:jwithdelta}
	\min_{\Delta x_\mathrm{i,k}} J(\Delta x_\mathrm{i,k}) = \frac{\eta_\mathrm{k}\|y_\mathrm{k} -  \sum_\mathrm{i=1}^\mathrm{q} w_\mathrm{i,k}^\mathrm{T}(x_\mathrm{FF,i,k}+\Delta x_\mathrm{i,k})\|^2}{2}
\end{equation} 
and calculate the gradient w.r.t. $\Delta x_\mathrm{i,k}$.
A key challenge is that infinitely many parameter combinations produce the same control output, so the adaptive controller can achieve correct performance without identifying the true parameter values. This could lead to parameter drift, where the sum $x_\mathrm{i,k} = x_\mathrm{FF,i,k} + \Delta x_\mathrm{i,k}$ diverges from the feedforward estimate, even if the control output remains correct. While the stability proof confirms that this does not lead to instability in the chosen setup, the benefits of improved performance during speed changes -- achieved through the delta learning approach -- could be compromised. To counteract this, a damping term is introduced into the adaptive law~\eqref{eq:al}, resulting in the modified adaptive law
\begin{equation} 
	\Delta x_\mathrm{i,k+1} = (1-\sigma)\Delta x_\mathrm{i,k} + \Gamma w_\mathrm{i,k} \epsilon_\mathrm{k},
\end{equation}
with the scalar design parameter $\sigma>0$. Note that this modification is known as $\sigma$-modification or leakage and is common to enhance robustness of adaptive control schemes \cite{robustAdaptiveControl,damping}. 

\subsection{Lookup table setup} \label{subsec:lut}

Initially, the \gls{LUT} is populated with identified parameters. Then, it is subsequently updated online. 

\subsubsection{Transfer function $G_\mathrm{u \rightarrow y}$}
The first two parameters in~\eqref{eq:estiparameters1} correspond to the real and imaginary parts of $G_\mathrm{u \rightarrow y}(j\omega_\mathrm{i})$, which vary with the speed $n_\mathrm{}$, but not with torque $T$. They are initialized offline by measuring at least two different \gls{HSS} responses and calculating the parameters via regression.

\subsubsection{Disturbance parameters $\theta_\mathrm{p,s}$ and $\theta_\mathrm{p,c}$}
The disturbance parameters depend on both $n_\mathrm{}$ and torque $T$ and are stored in a two-dimensional \gls{LUT}. They are initialized similarly from offline measurements at selected support points.

\noindent For both cases, feedforward parameter values at operating points between the support points are obtained via linear (for $G$) or bilinear (for $\theta_\mathrm{p}$) interpolation, and extrapolated if outside the grid. 

\subsection{Online-adaptation of the lookup table} \label{subsec:adaptiveLut}

Note that the scheme with the shifted optimization problem~\eqref{eq:jwithdelta} heavily relies on an accurate true parameter prediction. Wrong parameter estimates may even have a detrimental effect on the delta-learning approach. This is especially critical in e-drive systems, where the system behavior is inherently changing during its lifetime due to aging effects. Therefore, we propose an online-adaptation scheme of the \gls{LUT}.
\begin{figure}
	\centering
	\includegraphics[]{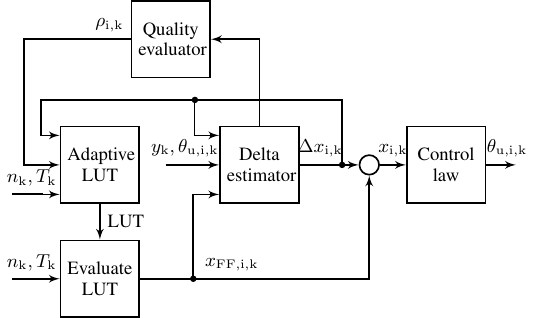}
	\caption{Adaptive delta learning structure.}
	\label{fig:adaptivelut}
\end{figure} 

Crucially, the \gls{LUT} shall be only updated, when (i) the feedforward parameters deviate from the actual estimated parameters $x_\mathrm{FF,i,k}+\Delta x_\mathrm{i,k}$, and (ii), if the data is informative enough. Point (i) is addressed by simply stopping any \gls{LUT}-updates when $\Delta x_\mathrm{i,k} = 0$. To achieve point (ii), we leverage the convergence rate $\rho_\mathrm{i,k}$ which is obtained from \autoref{prop:rhok}. Following the discussions of \autoref{sec:parameterconv}, if $\rho_\mathrm{i,k}<1$, then the data is suitable for parameter updates and a LUT update is desirable. The smaller $\rho_\mathrm{i,k}$ the better the regression equations are conditioned and the larger the \gls{LUT} update shall be. 

Consider \autoref{fig:adaptivelut}.
Compared to the structure shown in \autoref{fig:deltaadaptivecontroller}, two additional blocks are introduced.
The estimation quality is evaluated in the "Quality Evaluator" block, which leverages internal properties of the delta estimator to compute $\rho_\mathrm{i,k}$.

The \gls{LUT} is now generated by the new "Adaptive \gls{LUT}" block, which performs the \gls{LUT} update following
\begin{equation} \label{eq:adaptiveLUTupdate}
	\mathrm{LUT}(n_\mathrm{k},T_\mathrm{k}) \gets \mathrm{LUT}(n_\mathrm{k},T_\mathrm{k}) + (1 - \rho_\mathrm{i,k}) \, \beta \, \Delta x_\mathrm{i,k},
\end{equation}
where $\beta$ is the learning rate. 
The \gls{LUT} is continuously updated as long as $\rho_\mathrm{i,k} < 1$ and $\| \Delta x_\mathrm{i,k}\|> 0$. In other words, updates occur when the estimator is reliable and detects a discrepancy between the feedforward parameters and the true parameters.

\noindent\textbf{Remark.} 
We emphasize that dynamic regressor extension is only applied to compute $\rho_{\mathrm{i,k}}$ for the dynamic regressor extension and \emph{not} in the control law introduced in \autoref{sec:setup}. The main reason is the limited computational budget imposed by the typically short sampling times in e-drive systems. The LUT adaptation is not real-time critical and thus, dynamic regressor extension may be applied here.

\subsection{Active learning} \label{subsec:activelearning}

Parameter convergence requires sufficient excitation, i.e.\ $\rho_\mathrm{i,k}<1$. However, since the control law~\eqref{eq:control} enforces $y=0$, a lack of excitation will eventually be achieved by design. 
To address this, an active learning law is introduced. It is well known that for a $n$-th order system linear in its parameters, $\tfrac{n}{2}$ distinct frequencies in the reference signal are sufficient~\cite{boyd1}. 
Thus, consider the following excitation law
\begin{equation} \label{eq:alcontrol}
	\theta_\mathrm{u,i,k} = G_\mathrm{i,k}^{-1} \left(Y_\mathrm{des,i,k} - \theta_\mathrm{p,i,k} \right),
\end{equation}
with $Y_\mathrm{des,i,k}$ being an excitation signal
\begin{equation} \label{eq:ydessin}
	Y_\mathrm{des,i,k} = \begin{bmatrix}
		A_\mathrm{al1,i,k} \cos \left(\omega_\mathrm{al1,i} k t_\mathrm{s}\right) \\[3pt]
		A_\mathrm{al2,i,k} \cos \left(\omega_\mathrm{al2,i} k t_\mathrm{s}\right) 
	\end{bmatrix},
\end{equation}
containing two distinct frequencies $\omega_\mathrm{al1,i,k} \neq \omega_\mathrm{al2,i,k}$.
This enforces
\begin{equation} \label{eq:yal}
	y_\mathrm{k} = \sum_\mathrm{i=1}^\mathrm{q} f_\mathrm{i,k}^\mathrm{T} Y_\mathrm{des,i,k},
\end{equation}
which ensures sufficient excitation. 

Now note that permanent excitation is desirable for parameter convergence on the one hand, but undesirable on the performance level on the other hand, since $y$ still reflects an \gls{NVH} signal. Thus, active learning is activated only if $\Delta x_\mathrm{i,k}\neq 0$ and excitation is insufficient. The amplitudes are set as 
\begin{equation}
	A_\mathrm{al1,i,k} = A_\mathrm{al2,i,k} =
	\max{\left(0, \tfrac{\epsilon-(1-\rho_\mathrm{i,k})}{\epsilon}\right)}
	\, \delta \|\Delta x_\mathrm{i,k}\|,
\end{equation}
with tuning factor $\delta>0$ and threshold $\epsilon>0$. If $1-\rho_\mathrm{i,k} \geq \epsilon$, excitation is sufficient and active learning is disabled. 

Thus, the delta learning approach with adaptive \gls{LUT} ensures that $\Delta x_\mathrm{i,k} \to 0$, while active learning is only applied when necessary.
\section{Simulation results} \label{sec:simResults}

This section presents simulation results for the concepts introduced in \autoref{chap:concept}. The results are obtained from a closed-loop simulation in MATLAB/Simulink, which includes a simulation model of the \gls{PSM} and an implementation of a \gls{FOC} K to achieve the control loop presented in \autoref{fig:controlWoHHC}. 
The unknown NVH transfer behavior is modeled as a second order filter, which takes $i_\mathrm{q}$ as its input. Additionally, the disturbance $p$ is modeled as a sinusoidal signal in the \( 12^\mathrm{th} \) electrical harmonic frequency. This disturbance is added to the output of the NVH transfer function and yields \( y \). The \gls{HC} controllers are implemented and connected as described in \autoref{chap:concept}. To enhance realism, random noise is added to the rotational speed, the \gls{PSM} currents, and the measured NVH signal \( y \). The amplitudes of the disturbance and the random noise are chosen to match the levels observed on the testbench, which is introduced in the next section.

\begin{figure}[!t]
	\centering
	\includegraphics[width=0.5\textwidth]{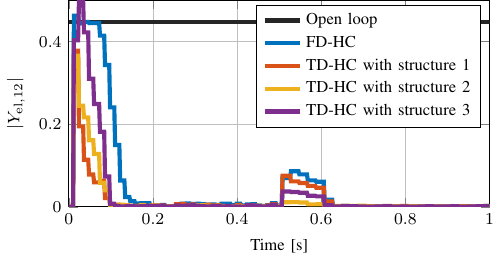}
	\caption{Reduction of the $12^\mathrm{th}$ electrical harmonic in $y$ comparing the time-domain control structures with the benchmark controller.}
	\label{fig:sim2}
\end{figure}

In \autoref{fig:sim2}, the results are shown for all three control structures, along with the comparison of an adaptive frequency-domain HC controller \cite{kamaldarFHC} as benchmark.
These results are derived from a simulation where the rotational speed is set to \( n_\mathrm{mech} = 1000 \ \mathrm{rpm} \) for the first \( \SI{0.5}{\second} \) and then reduced to \( n_\mathrm{mech} = 800 \ \mathrm{rpm} \) for the remaining \( \SI{0.5}{\second} \). The presented signals for $|Y_\mathrm{el,12}|$ -- the amplitudes of the \( 12^\mathrm{th} \) electrical harmonic in \( y \) -- are the result of a \gls{DFT} analysis performed at the end of each electrical period, which leads to the discrete update steps of the signal.
For each control structure, the gain \( \gamma \) is selected through a basic parameter tuning process to ensure that each structure performs well individually.
\autoref{tab:stats} evaluates the results based on five performance indicators related to convergence speed and steady-state error. 
\begin{table}
	\centering
	\caption{Performance indicators for different control structures.}
	\begin{tabular}{l cccc} 
		\toprule
		\textbf{Performance Indicator} & \textbf{FD} & \textbf{TD S1} & \textbf{TD S2} & \textbf{TD S3} \\ 
		\midrule
		Time until $|Y_\mathrm{el,12}|$ is $\leq 0.05$ & \cellcolor{red!50}0.132 & \cellcolor{green!50}0.084 & \cellcolor{green!50}0.084 & \cellcolor{green!30}0.096 \\
		Mean of $|Y_\mathrm{el,12}|$& \cellcolor{red!50}0.052 & \cellcolor{green!40}0.019 & \cellcolor{green!50}0.017 & \cellcolor{green!20}0.03 \\
		Maximum $|Y_\mathrm{el,12}|$ after 0.5s & \cellcolor{red!50}0.086 & \cellcolor{red!30}0.076 & \cellcolor{green!50}0.012 & \cellcolor{green!30}0.037 \\
		Mean of $|Y_\mathrm{el,12}|$ in Interval 1 & \cellcolor{red!50}0.005 & \cellcolor{green!20}0.003 & \cellcolor{green!40}0.002 & \cellcolor{green!50}0.002 \\
		Mean of $|Y_\mathrm{el,12}|$ in Interval 2 & \cellcolor{green!30}0.001 & \cellcolor{green!30}0.001 & \cellcolor{green!30}0.001 & \cellcolor{green!50}0.001 \\ 
		\bottomrule
	\end{tabular}
	\label{tab:stats}
	\begin{tablenotes}
		\small
		\item Interval 1: $t=[\SI{0.2}{\second}, \SI{0.5}{\second}]$, Interval 2: $t=[\SI{0.7}{\second}, \SI{1.0}{\second}]$, FD: frequency-domain, TD: time-domain
	\end{tablenotes}
\end{table}

Among the three time-domain control structures, no major differences in performance are observed; variations due to different gains or initial conditions are larger than differences between the structures themselves. All three, however, outperform the benchmark controller in terms of faster convergence and lower mean values of $|Y_{\text{el},12}|$.
\section{Testbench results} \label{sec:testbenchResults}

In this section, the adaptive time-domain \gls{HC}, along with its extensions, is examined on the testbench. In all presented results, the rotational speed $n_\mathrm{mech}$, the torque $T$ and the measured laser signal $y$ are given normalized by scaling factors.

\subsection{Testbench setup} \label{sec:experiment}

\begin{figure}
	\centering
	\includegraphics[width=0.35\textwidth]{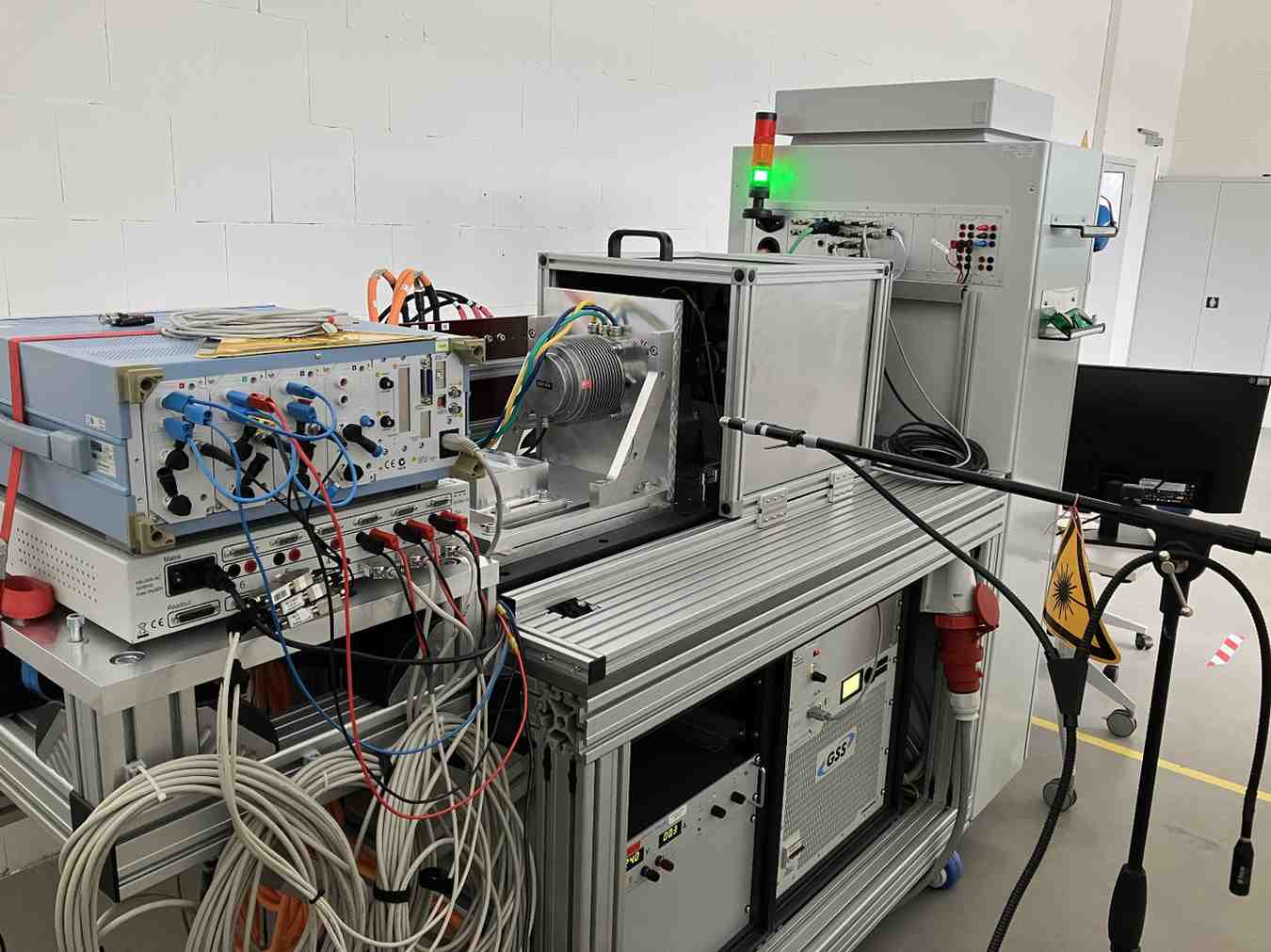}
	\caption{Testbench setup.}
	\label{fig:setup}
\end{figure}

The testbench setup is shown in \autoref{fig:setup}. It shows the \gls{LDV} measuring \gls{NVH} emissions in the form of surface vibrations on the housing of the used 48V \gls{PSM}.
The \gls{PSM} is coupled to a load machine, which regulates the rotational speed to maintain a constant reference value.
The controller is implemented in MATLAB/Simulink and online computation is performed on a dSPACE RTI 1007 setup. The sampling rate is set to \SI{10}{\kilo \hertz}, corresponding to a sampling time of $t_\mathrm{s} = \SI{100}{\micro \second}$.
To evaluate \gls{NVH} reduction under different conditions, various operating points are tested in the experiment to assess the performance of the controllers.
This experiment is well-suited for controller evaluation, as all parameters estimated online vary throughout the experiment.
Here, mainly the results of reducing the $12^\mathrm{th}$ electrical harmonic are presented. In \autoref{sec:multiFreq}, it is demonstrated how the approach can be applied to reduce multiple harmonics. 

\subsection{Open loop experiment}

\begin{figure}[!t]
	\centering
	\includegraphics[width=0.5\textwidth]{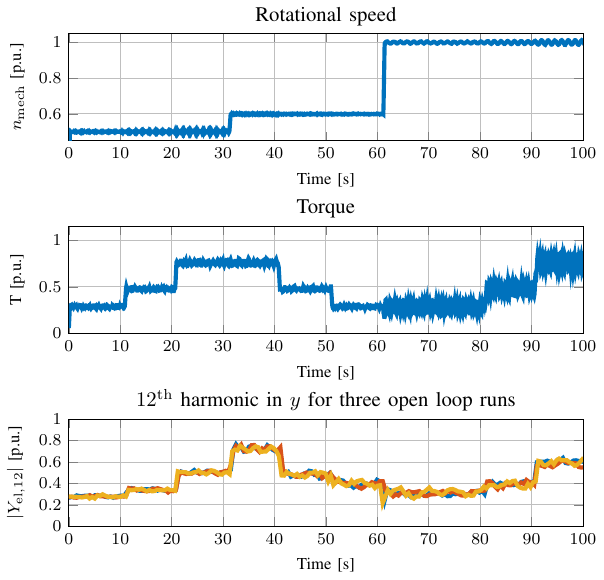}
	\caption{Open loop experiment: Rotational speed, torque and the amplitude of the $12^\mathrm{th}$ electrical harmonic in the laser signal.}
	\label{fig:doe}
\end{figure}

\begin{figure}[!t]
	\centering
	\includegraphics[width=0.5\textwidth]{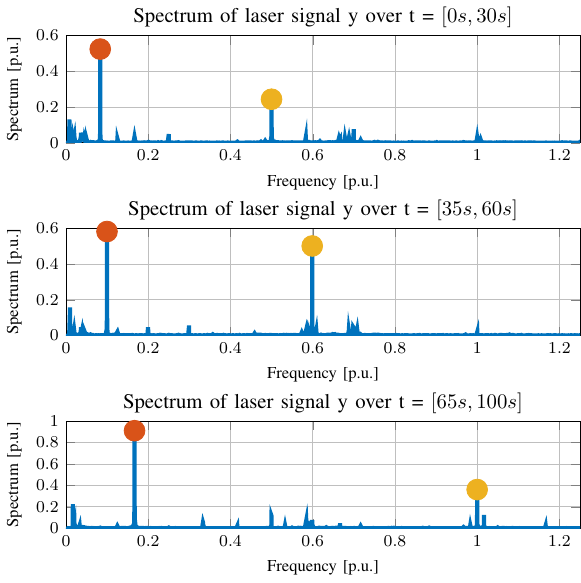}
	\caption{Frequency spectrum of the laser signal over the different constant speed intervals from an open loop testbench experiment. Blue shows the spectrum; red and yellow markers highlight the $2^\mathrm{nd}$ and $12^\mathrm{th}$ harmonics, respectively.}
	\label{fig:spectrum}
\end{figure}

To demonstrate the nominal \gls{NVH} pollution, an open loop experiment without \gls{HC} is performed.
The results of this experiment using the control loop without \gls{HC} are shown in \autoref{fig:doe}. 
The third subplot shows the amplitude of the laser signal $y$ in the $12^\mathrm{th}$ electrical frequency for three runs of the same experiment.
The complete frequency spectrum of $y$ over all three rotational speeds is shown in \autoref{fig:spectrum}. In all three intervals, the most prominent frequency components are the $2^\mathrm{nd}$ and $12^\mathrm{th}$ electrical harmonics.

\subsection{Adaptive time-domain harmonic control} \label{sec:testbenchResultsConcept}

\begin{figure}[!t]
	\centering
	\includegraphics[width=0.5\textwidth]{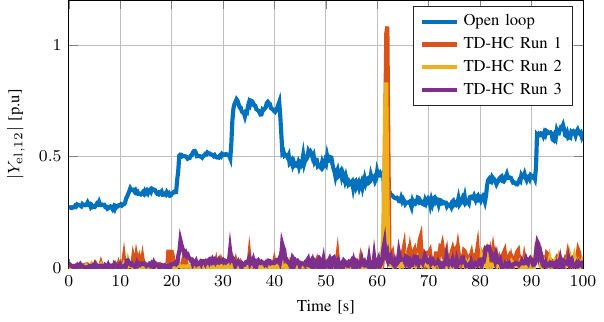}
	\caption{Reduction of the $12^\mathrm{th}$ electrical harmonic in the laser signal.}
	\label{fig:td_hhc}
\end{figure}

Here, the experiment is repeated with the adaptive time-domain \gls{HC} deployed. \autoref{fig:td_hhc} shows the outcomes of three runs of the same experiment. Due to space reasons, only the results for structure 1 are displayed, however, we note that the experiment outcomes do only differ slightly. In fact, the discrepancy among different runs for the same control structure -- originating from different parameter initializations -- is larger than the differences among the structures themselves. Thus, the insights discussed here for Structure 1 also apply to the other two structures.
The $12^\mathrm{th}$ electrical harmonic is significantly reduced for most of the duration, while the performance after the speed change from $0.6 \ \mathrm{p.u.}$ to $1 \ \mathrm{p.u.}$ varies considerably between runs. For the different runs, the same controller with the same parameterization is used, with the initial parameter values being the only difference.

\begin{figure}[!t]
	\centering
	\includegraphics[width=0.5\textwidth]{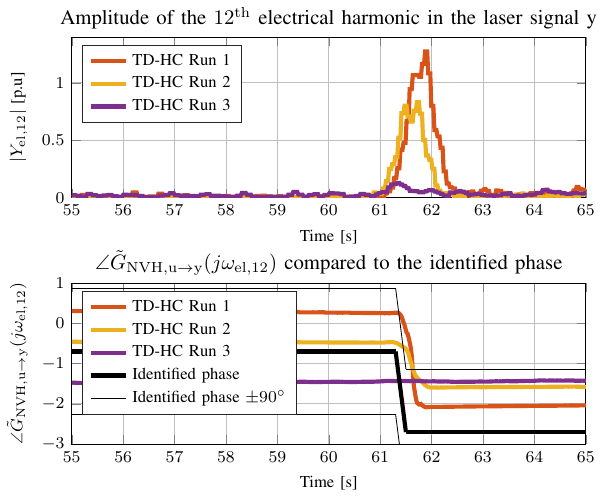}
	\caption{Amplitude $|Y_\mathrm{el,12}|$ and the estimated phase of $G_\mathrm{NVH,u \rightarrow y}(j\omega_\mathrm{el,12})$ compared to the identified phase during the speed change.}
	\label{fig:ampAndPhase}
\end{figure}

To explain the spikes around $60$ seconds, and why only run 1 and 2 are affected, observe \autoref{fig:ampAndPhase} that shows the identified and estimated phase $\angle \hat{G}_\mathrm{u\rightarrow y}(j \omega)$. Recall that outside a phase range of $|\angle G^*_\mathrm{u\rightarrow y}(j \omega) - \angle \hat{G}_\mathrm{u\rightarrow y}(j \omega)| < 90^{\circ}$, the nominal \gls{HC} becomes unstable. We indeed observe that our adaptive formulation can leave this range without harming stability, but with detrimental effects on the transient performance. In contrast, observe how run 3 stays within the nominal stable range and thus, shows only a mild spike in the amplitude. 

\subsection{Delta learning approach} \label{sec:deltaResults}

\begin{figure}[!t]
	\centering
	\includegraphics[width=0.5\textwidth]{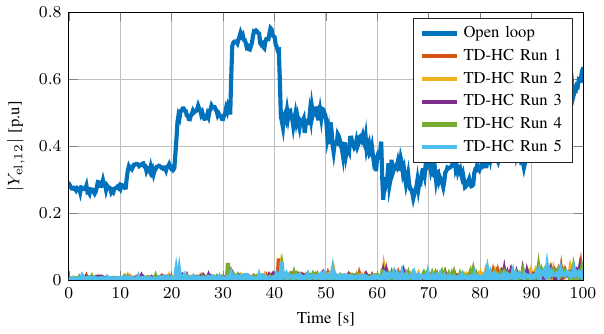}
	\caption{Reduction of the $12^\mathrm{th}$ electrical harmonic in the laser signal using control Structure 1 with the delta learning approach.}
	\label{fig:deltaResults}
\end{figure}

Note that the observations in \autoref{fig:ampAndPhase} motivate the use of the identified parameters as feedforward estimates in the proposed delta learning scheme described in \autoref{sec:deltaLearning}. The results of five experimental runs are illustrated in \autoref{fig:deltaResults}. 
The results indicate a significant reduction in the $12^\mathrm{th}$ electrical harmonic across all experiment runs throughout the complete duration of the experiment. 

\begin{figure}[!t]
	\centering
	\includegraphics[width=0.5\textwidth]{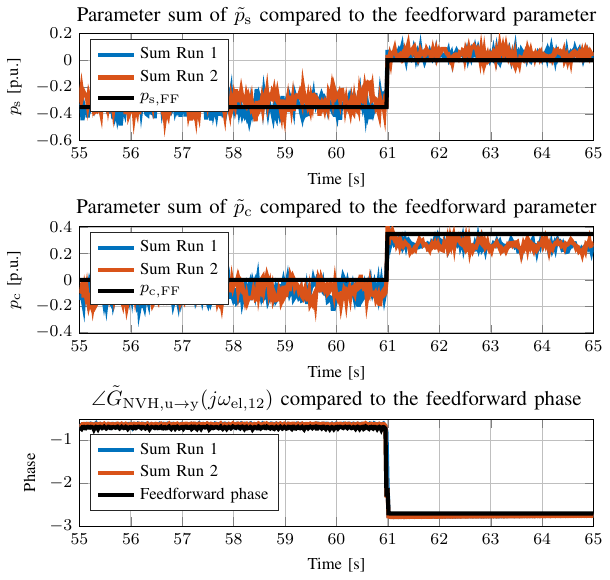}
	\caption{Parameter sum used in the controller compared to feedforward parameter value using control Structure 1 with the delta learning approach.}
	\label{fig:deltaParams}
\end{figure}

Crucially, observe how the spike from \autoref{fig:td_hhc} can be successfully mitigated, highlighting the effect of the delta learning approach. This complies with \autoref{fig:deltaParams}, where again the phase difference of identified and estimated system is displayed. Observe in the third subplot how the estimated phase, consisting of prior information plus delta estimation, now aligns well with the identified phase of the system, in contrast to \autoref{fig:ampAndPhase}. Moreover, observe in the first two subplots how incorrect feedforward parameters are counteracted by the delta estimate.
In particular, the delta estimator captures small high-frequency changes in the parameters. 

\subsection{Online-adaptation of the lookup table}

\begin{figure}[!t]
	\centering
	\includegraphics[width=0.5\textwidth]{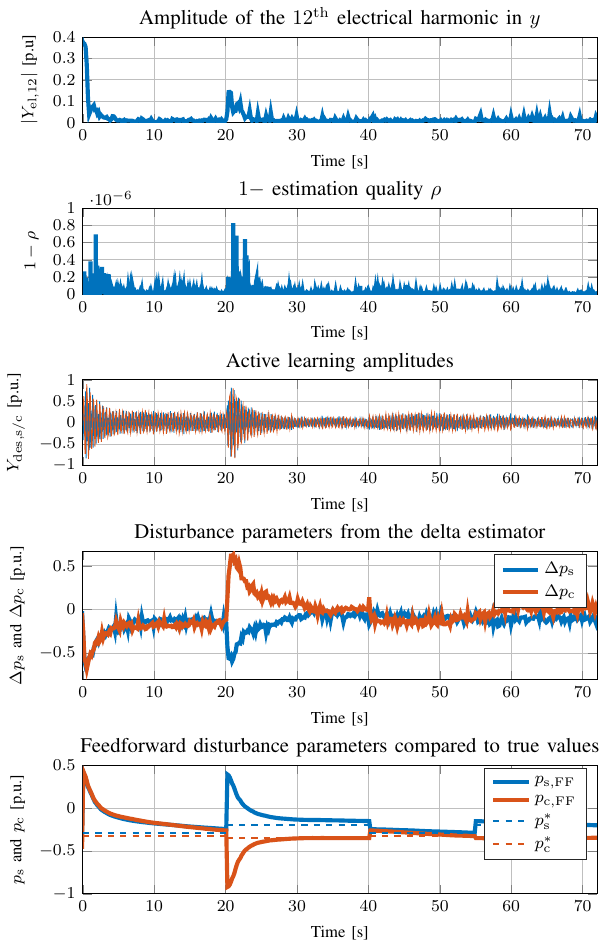}
	\caption{Reduction of the $12^\mathrm{th}$ electrical harmonic in the laser signal using the delta learning approach with online-adaptation of the lookup table.}
	\label{fig:activelearningdeltatestbench}
\end{figure}

Here, testbench results for the online adaptation of the \gls{LUT} from \autoref{subsec:adaptiveLut} in combination with active learning from \autoref{subsec:activelearning} are presented. In \autoref{fig:activelearningdeltatestbench}, results are shown for a testbench experiment operated at two operating points with different torques, each tested twice. Initially, $\theta_\mathrm{p,s,FF}$ and $\theta_\mathrm{p,c,FF}$ show a large deviation from the true parameters $\theta_\mathrm{p,s}^*$ and $\theta_\mathrm{p,c}^*$ due to incorrect \gls{LUT} entries.
The results demonstrate that when an operating point with significant deviation is reached, the delta estimator estimates the difference between the feedforward and true parameters. This is achieved by enabling active learning which leads to a high estimation quality. The \gls{LUT} is updated, reducing the deviation between the feedforward and true parameters, which in turn lowers the delta parameters.
When the operating point is reached a second time, performance in reducing the $12^\mathrm{th}$ electrical harmonic improves significantly, as shown by the performance indicators in \autoref{tab:statstb2}. 

\begin{table}[h]
	\centering
	\caption{Performance indicators for reaching an operating point with incorrect feedforward estimates and a second time with updated values.}
	\begin{tabular}{l cc} 
		\toprule
		\textbf{Performance Indicator} & \textbf{First time} & \textbf{Second time} \\ 
		\midrule
		Maximum of $|Y_\mathrm{el,12}|$ for $T = 0.5 \ \mathrm{p.u.}$ & 0.4068 & 0.0314 \\
		Maximum of $|Y_\mathrm{el,12}|$ for $T = 0.3 \ \mathrm{p.u.}$ & 0.1436 & 0.0436 \\
		Mean value of $|Y_\mathrm{el,12}|$ for $T = 0.5 \ \mathrm{p.u.}$ & 0.0267 & 0.0097 \\
		Mean value of $|Y_\mathrm{el,12}|$ for $T = 0.3 \ \mathrm{p.u.}$ & 0.0202 & 0.0116 \\
		\bottomrule
	\end{tabular}
	\label{tab:statstb2}
\end{table}

\subsection{Reduction of multiple frequencies} \label{sec:multiFreq}

\begin{figure}[!t]
	\centering
	\includegraphics[width=0.5\textwidth]{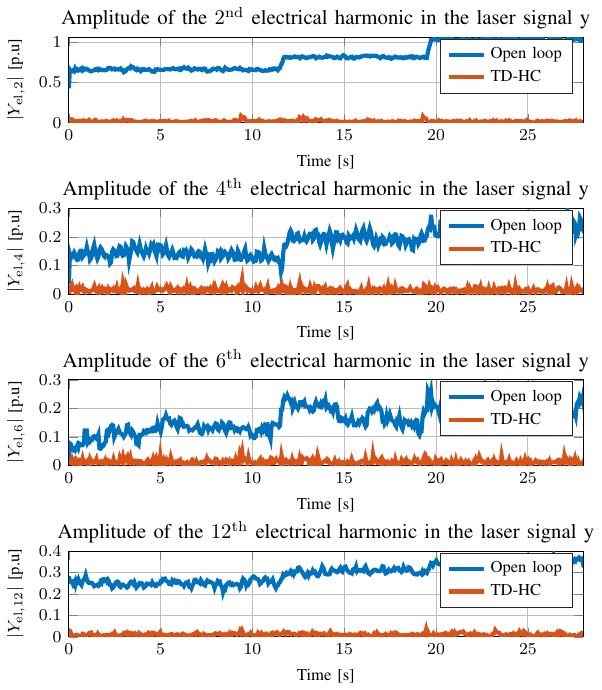}
	\caption{Simultaneous reduction of multiple electrical harmonics.}
	\label{fig:multiFreqResults}
\end{figure}

\autoref{fig:multiFreqResults} illustrates the results for reducing four different disturbance frequencies simultaneously for an experiment where the operating point is changed twice. It is evident that the $2^\mathrm{nd}$, $4^\mathrm{th}$, $6^\mathrm{th}$ and $12^\mathrm{th}$ electrical harmonics can be successfully reduced in parallel.
\section{Conclusion} \label{sec:conclusion}

This work advances adaptive time-domain \gls{HC} from theory to practice for \gls{NVH} mitigation in electric drives. Three control structures were developed to integrate the adaptive controller into existing loops, and the estimation scheme was refined to reduce computational effort for embedded use. 

Simulations and testbench experiments confirm faster convergence than frequency-domain \gls{HC}, but also reveal limitations during rapid speed changes. To ensure robustness across operating points, a delta learning scheme was introduced that combines the adaptive controller with a \gls{LUT}-based feedforward estimator, complemented by an adaptive \gls{LUT} for lifetime robustness through online updates.

Future work could include a rigorous stability analysis of the extended delta learning structure to ensure that the combination of the adaptive controller and adaptive \gls{LUT} does not introduce undesired interactions in the overall system. In addition, while the adaptive time-domain \gls{HC} with delta learning has shown promising performance, other adaptive control concepts recently proposed in the literature -- such as sliding mode control \cite{outlook5} or predictive cost adaptive control \cite{outlook7} -- could be explored.
{\appendices
	\section{Proof sketch of \autoref{stabilityproof}} \label{proof}
	\begin{proof}
		Subtracting $\alpha x_\mathrm{i}^*$ from both sides of~\eqref{eq:al} yields 
		\begin{equation} \label{eq:xesti}
			\tilde{x}_\mathrm{i,k+1} = \tilde{x}_\mathrm{i,k} + \Gamma w_\mathrm{i,k} \epsilon_\mathrm{k}
		\end{equation}
		Considering the individual estimation errors of~\eqref{eq:estierror} as
		\begin{align} \label{eq:estierrorHp}
			\tilde{G}_\mathrm{i,k} &= G_\mathrm{i,k} - \alpha G_\mathrm{i}^* \quad & \tilde{\theta}_\mathrm{p,i,k} &= \theta_\mathrm{p,i,k} - \alpha \theta_\mathrm{p,i}^*,
		\end{align}
		and splitting~\eqref{eq:xesti} yields
		\begin{subequations} \label{eq:tildeHeasy}
			\begin{equation} 
				\tilde{G}_\mathrm{i,k+1} = \tilde{G}_\mathrm{i,k} + \gamma_\mathrm{G} \eta_\mathrm{k} W_\mathrm{G,i,k} \bar{y}_\mathrm{k}
			\end{equation} 
			with
			\begin{equation}
				W_\mathrm{G,i,k} =
				\begin{bmatrix}
					a & -b \\ b & a
				\end{bmatrix},
				\quad
				\begin{aligned}
					a &= \theta_\mathrm{u,s,i,k} s_\mathrm{i,k} + \theta_\mathrm{u,c,i,k} c_\mathrm{i,k} \\
					b &= \theta_\mathrm{u,s,i,k} c_\mathrm{i,k} - \theta_\mathrm{u,c,i,k} s_\mathrm{i,k}
				\end{aligned}
			\end{equation}
			and
			\begin{equation} 
				\tilde{\theta}_\mathrm{p,i,k+1} = \tilde{\theta}_\mathrm{p,i,k} + \gamma_\mathrm{p} \eta_\mathrm{k} \begin{bmatrix} s_\mathrm{i,k} \\[2pt] c_\mathrm{i,k} \end{bmatrix} \bar{y}_\mathrm{k}
			\end{equation}
			for $s_\mathrm{i,k} = \sin\left(\omega_\mathrm{i} k t_\mathrm{s}\right)$ and $c_\mathrm{i,k} = \cos\left(\omega_\mathrm{i} k t_\mathrm{s}\right)$.
		\end{subequations}	
		Defining the Lyapunov function measuring all parameter estimation errors
		\begin{equation} \label{eq:V}
			V(\tilde{x}_\mathrm{1,k} \dots \tilde{x}_\mathrm{q,k}) = \frac{2}{\gamma_\mathrm{p}}\sum_\mathrm{i=1}^\mathrm{q} \|\tilde{\theta}_\mathrm{p,i,k}\|^2 + \frac{1}{\gamma_\mathrm{G}} \sum_\mathrm{i=1}^\mathrm{q} \|\tilde{G}_\mathrm{i,k}\|_\mathrm{F}^2
		\end{equation}
		the corresponding Lyapunov difference is
		\begin{equation}
			\Delta V(k) = V\left(\tilde{x}_\mathrm{1,k+1} \dots \tilde{x}_\mathrm{q,k+1}\right) - V\left(\tilde{x}_\mathrm{1,k} \dots \tilde{x}_\mathrm{q,k}\right).
		\end{equation}
		Using straightforward reformulations and noting from~\eqref{eq:normalization} that $\eta_\mathrm{k} \leq 1$ and $\eta_\mathrm{k} \sum_\mathrm{i=1}^\mathrm{q} \|\theta_\mathrm{u,i,k}\|^2 \leq 1$,
		one obtains the bound
		\begin{equation}
			\Delta V(k) \leq -4 \eta_\mathrm{k} \left( \alpha - q \gamma_\mathrm{p} - \frac{\gamma_\mathrm{G}}{2} \right) y_\mathrm{k}^2.
		\end{equation}
		Since this is the same bound as in~\cite{kamaldarTHC}, all claims of \autoref{stabilityproof} follow by the standard arguments in~\cite{kamaldarTHC}.
	\end{proof}

\bibliographystyle{ieeetr}
\bibliography{root}

\begin{IEEEbiography}[{\includegraphics[width=1in,height=1.25in,clip,keepaspectratio]{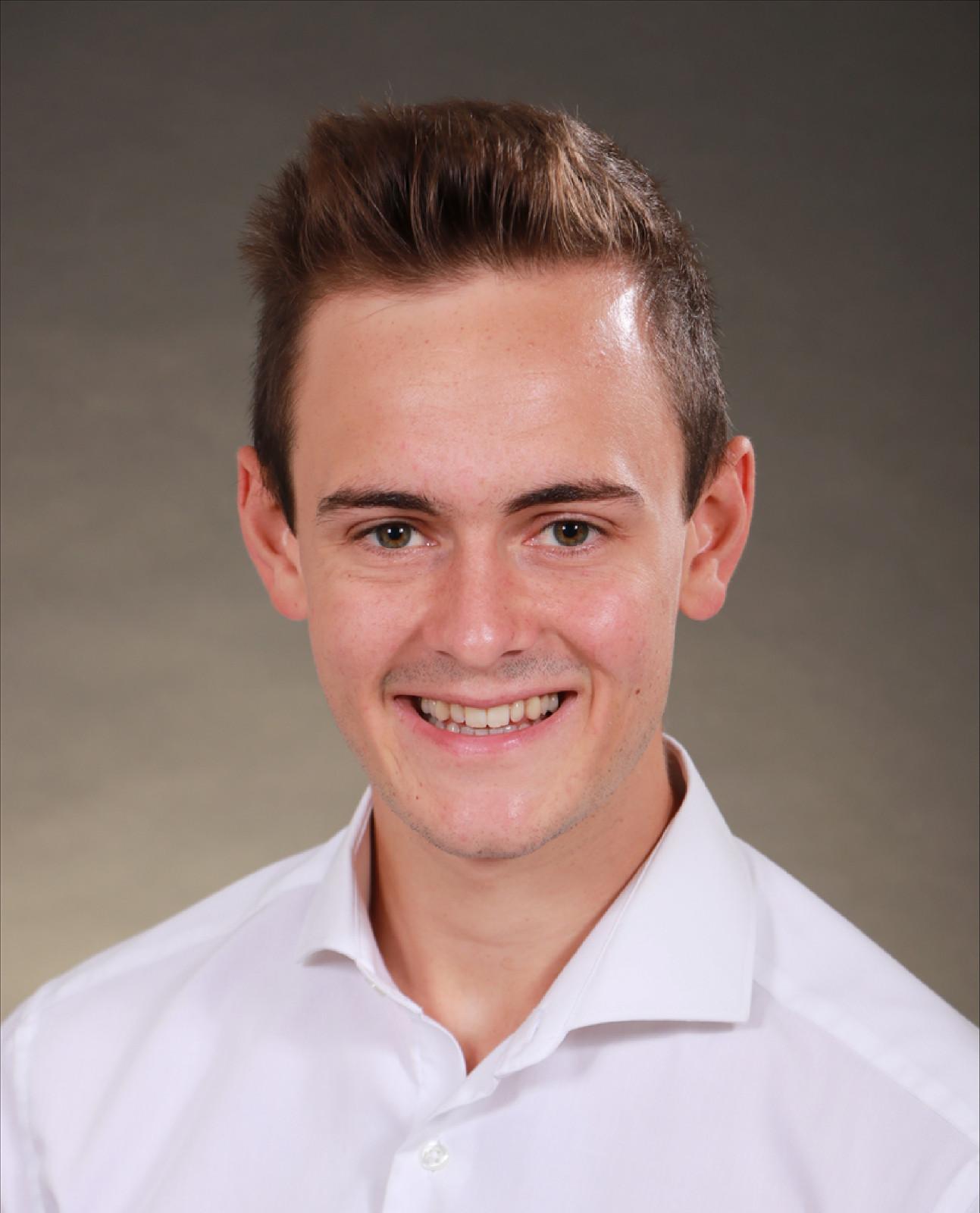}}]{Klaus Herburger}
	received the B.Sc. and M.Sc. degrees in engineering cybernetics from the University of Stuttgart, Stuttgart, Germany, in 2021 and 2024, respectively. 
	He conducted his master’s thesis at Bosch Research, Renningen, Germany, in cooperation with the Institute for Systems Theory and Automatic Control, University of Stuttgart. 
	Since 2022, he has been with Robert Bosch GmbH, Leonberg, as an engineer in the Advanced Driver Assistance Systems domain.
\end{IEEEbiography}

\begin{IEEEbiography}[{\includegraphics[width=1in,height=1.25in,clip,keepaspectratio]{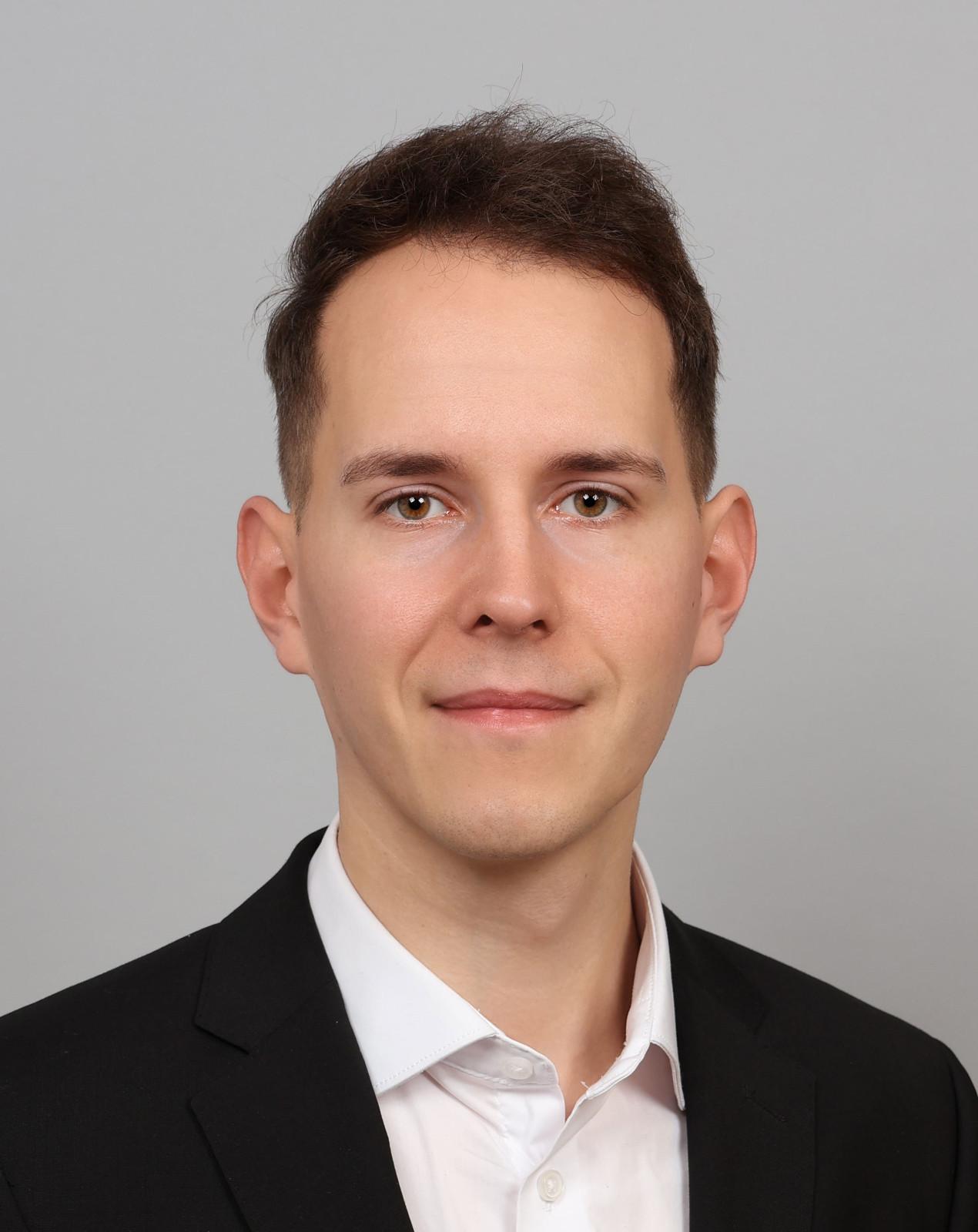}}]{Fabian Jakob}
	received the B.Sc. and M.Sc.
	degrees in engineering cybernetics from the University of Stuttgart, Stuttgart, Germany, in 2020 and 2022, respectively. From 2021 to 2022, he worked as a research intern in the electric drive department of Bosch Research. He is currently pursuing the Ph.D. degree at the Institute for Systems Theory and Automatic Control, University of Stuttgart, under the supervision of Prof. Dr. Andrea Iannelli. His research interests are the exploration of robust and adaptive control methods for optimization algorithms.
\end{IEEEbiography}

\begin{IEEEbiography}[{\includegraphics[width=1in,height=1.25in,clip,keepaspectratio]{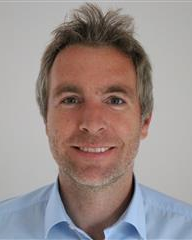}}]{David G\"anzle}
	received his Diploma degree in mechanical engineering from the University of Stuttgart, Germany, in 2007. He joined Robert Bosch GmbH in the same year and is currently working at Bosch Research in Renningen, Germany. His work focuses on the modeling, combination of physics-based and data-driven approaches, and control of electrical drives in heating systems and automotive applications. His research interests include adaptive control, such as noise, vibration, and harshness reduction for electric drives.
\end{IEEEbiography}

\begin{IEEEbiography}[{\includegraphics[width=1in,height=1.25in,clip,keepaspectratio]{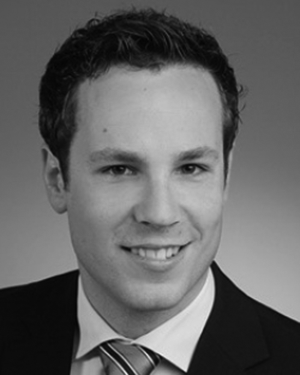}}]{Maximilian Manderla}
	received the Diploma degree in mechanical engineering and the Dr.-Ing. degree (Ph.D.) in control systems engineering from the Technical University of Darmstadt, Germany, in 2007 and 2011, respectively. He studied mechanical engineering from the Technical University of Darmstadt and the University of California at Berkeley. He was working with Voith Hydro, Heidenheim, Germany, from 2011 to 2015 with focus on hydro-electric power plant dynamics, simulation, and control. In 2015, he joined Bosch, Renningen, Germany, where he is currently active as a Project Manager in the field of control engineering with emphasis on electrical drives in automotive applications. His research interests include model-predictive and learning-based control.\end{IEEEbiography}

\begin{IEEEbiography}[{\includegraphics[width=1in,height=1.25in,clip,keepaspectratio]{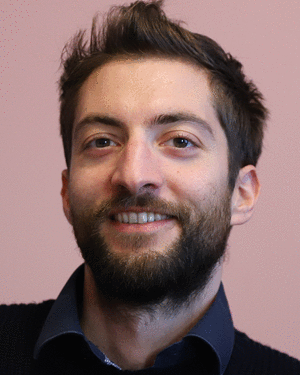}}]{Andrea Iannelli}
	(Member, IEEE) is an Assistant Professor in the Institute for Systems Theory and Automatic Control at the University of Stuttgart. He completed his B.Sc. and M.Sc. degrees in Aerospace Engineering at the University of Pisa and received his PhD from the University of Bristol. He was also a postdoctoral researcher in the Automatic Control Laboratory at ETH Zurich. His main research interests are centered around robust and adaptive control, uncertainty quantification, and sequential decision-making. He serves the community as Associated Editor for the International Journal of Robust and Nonlinear Control and as IPC member of international conferences in the areas of control, optimization, and learning.\end{IEEEbiography}

\end{document}